\documentstyle[prl,aps,latexsym,graphicx,twocolumn]{revtex}
\begin{document}
\draft
\title{Strong ``quantum'' chaos in the global ballooning 
mode spectrum of three-dimensional plasmas}
\author{R. L. Dewar\thanks{Permanent address: Research School of 
Physical Sciences \& Engineering, The Australian National 
University. E-mail: robert.dewar@anu.edu.au.}}
\address{Princeton University Plasma Physics Laboratory, P.O. Box 
451, Princeton N.J. 08543}
\author{P. Cuthbert and R. Ball}
\address{Department of Theoretical Physics and Plasma Research Laboratory, 
Research School of Physical Sciences \& Engineering,
The Australian National University,
Canberra 0200 Australia}
\date{\today}
\maketitle

\newcommand{\eq}[1]{Eq.~(\ref{#1})}
\newcommand{\fig}[1]{Fig.~\ref{#1}}         
\renewcommand{\d}{{\,\mathrm{d}}} % differential
\newcommand{\kappav}{\mbox{\boldmath\(\kappa\)}}
\newcommand{\kappan}{{\kappa_{\mathrm{n}}}}
\newcommand{\kappag}{{\kappa_{\mathrm{g}}}}
\newcommand{\dotv}{\mbox{\boldmath\(\cdot\)}}
\newcommand{\cross}{\mbox{\boldmath\(\times\)} }
\newcommand{\grad}{\mbox{\boldmath\(\nabla\)} }
\newcommand{\divv}{\mbox{\boldmath\(\nabla\cdot\)} }
\newcommand{\curl}{\mbox{\boldmath\(\nabla\times\)} }
\newcommand{\const}{{\mathrm{const}}}
\newcommand{\Bvec}{{\mathbf{B}}}
\newcommand{\kvec}{{\mathbf{k}}}
\newcommand{\AGC}{{\mathcal{A}}}
\newcommand{\KGC}{{\mathcal{K}}}
\newcommand{\NGC}{{\mathcal{N}}}
\newcommand{\omegaGC}{{\mathcal{N}}\omega^2 \rho}
\newcommand{\irs}{{\mathcal{R}}}

\bibliographystyle{prsty}

\begin{abstract}
	The spectrum of ideal magnetohydrodynamic (MHD) pressure-driven
	(ballooning) modes in strongly nonaxisymmetric toroidal systems is
	difficult to analyze numerically owing to the singular nature of
	ideal MHD caused by lack of an inherent scale length.  In this
	paper, ideal MHD is regularized by using a $k$-space cutoff,
	making the ray tracing for the WKB ballooning formalism a chaotic
	Hamiltonian billiard problem.  The minimum width of the toroidal
	Fourier spectrum needed for resolving toroidally localized
	ballooning modes with a global eigenvalue code is estimated from
	the Weyl formula.  This phase-space-volume estimation method is
	applied to two stellarator cases.
\end{abstract}

In design studies for new magnetic confinement devices for fusion
plasma experiments (e.g. investigations
\cite{reiman_etal97,reiman_etal99} leading to the proposed National
Compact Stellarator Experiment, NCSX \cite{nielson_etal00}), the
maximum pressure that can stably be confined in any proposed magnetic
field configuration is routinely estimated by treating the plasma as
an ideal magnetohydrodynamic (MHD) fluid.  One linearizes about a
sequence of equilibrium states with increasing pressure, and studies
the spectrum of normal modes (frequency $\omega$) to determine when
there is a component with $\mathrm{Im}\,\omega > 0$, signifying
instability.

Even with the simplification obtained by using the ideal MHD model,
the computational task of determining the theoretical stability of a
three-dimensional (i.e. nonaxisymmetric) device, such as NCSX or the
four currently operating helical axis stellators \cite{blackwell01},
remains a challenging one.

The problem can be posed as a Lagrangian field theory, with the
potential term being the energy functional $\delta W$
\cite{bernsteinetal58}.  For a static equilibrium, the kinetic energy
is quadratic in $\omega$, so that $\omega^2$ is real.  Thus
instability occurs when $\omega^2<0$.  There are two main approaches
to analyzing the spectrum---local and global.

In the \emph{local} approach, which is used for analytical
simplification, one orders the scale length of variation of the
eigenfunction across the magnetic field lines to be short
compared with equilibrium scale lengths \cite{dewar97c}.
Both interchange and ballooning stability can be treated by solving the
general \emph{ballooning equations} \cite{dewar-glasser83}, a system of
ordinary differential equations defined on a given magnetic field
line.

The \emph{global} (Galerkin) approach is to expand the plasma
displacement field in a finite basis set, inserting this ansatz in the
Lagrangian to find a matrix eigenvalue representation of the spectral
problem.  This approach has been implemented for ideal MHD in
three-dimensional plasmas in two codes, TERPSICHORE
\cite{anderson_etal90} and CAS3D \cite{schwab93}.

Although the Galerkin approach is potentially exact, if one could use
a complete, infinite basis set, it is in practice computationally
challenging due to the large number of basis functions required to
resolve localized instabilities.  This leads to very large matrices
which must be diagonalized by iterative methods.  There is a need for
analytical insight to determine a suitable truncated basis set and to
predict the nature of the spectrum, e.g. whether it is continuous or
discrete.

Such insight may be obtained by a \emph{hybrid} local-global approach,
in which one uses a Wentzel--Kramers--Brillouin (WKB) representation
of the eigenfunction.  In the short-wavelength limit, the same
analytical simplifications as are obtained in the local
approach are found to give a local dispersion relation that can be
used to give information on the global spectrum by using ray tracing
and semiclassical quantization.

In axisymmetric systems \cite{dewar-manickam-grimm-chance81} or in
cases where helical ripple can be averaged out, giving an adiabatic
invariant, \cite{cooper-singleton-dewar96,cuthbert_etal98}, the ray
equations are integrable and hence the spectrum is characterized by
``good quantum numbers''.

However, it has been known for many years \cite{dewar-glasser83} that
the ray-tracing problem in strongly three-dimensional systems is
singular because, in the absence of an adiabatic invariant, the
phase-space motion is not bounded---the rays escape to infinity in the
wavevector sector.  Dewar and Glasser \cite{dewar-glasser83} argued
that this gives rise to a \emph{continuous} unstable spectrum, with
correspondingly singular generalized eigenfunctions.  (A more rigorous
treatment involves the concept of the essential spectrum and Weyl
sequences \cite{hameiri85,lifschitz89}.)

Our proposed regularization of this singularity can be understood
using a simple quantum analogy.  Consider the one-dimensional
time-independent Schr\"{o}dinger equation $H\psi = E\psi$ in the limit
as the mass of the particle goes to infinity.  Then the kinetic energy
disappears and the Hamiltonian becomes $H = V(x)$, where $V$ is the
potential energy, assumed here to be the harmonic oscillator
potential, $\frac{1}{2}x^2$ in suitable units.  In the usual Hilbert
space the energy spectrum is continuous: $E \ge 0$ and the
(generalized) eigenfunctions singular: $\psi(x) =
\delta(x-x_{\mathrm{E}}) \pm \delta(x+x_{\mathrm{E}})$, where
$V(x_{\mathrm{E}}) \equiv E$.

We now seek a regularization of this problem by restricting $\psi$ to 
the space of functions with a finite bandwidth in wavenumber $k$:
\begin{equation}
	\psi(x) = \int^{k_{\max}}_{-k_{\max}}\frac{dk}{2\pi}\,
	\psi_k \exp ikx	\;.
	\label{eq:ktrunc}
\end{equation}
This truncated Fourier-integral representation models what occurs when
one seeks to find the spectrum numerically using a truncated
Fourier-series representation.

We take as starting point a Lagrangian for the wavefunction, 
\begin{equation}
	L = \int_{-\infty}^{\infty} \psi^{*}[E - V(x)]\psi \, dx \;.
	\label{eq:LSchr}
\end{equation}
Inserting \eq{eq:ktrunc} in \eq{eq:LSchr} gives
\begin{eqnarray}
	L & = & \int^{k_{\max}+0}_{-k_{\max}-0}
	\left[
		 E|\psi_k|^2 -
		 \left|
		 \frac{d\psi_k}{dk}+\psi_k\delta(k+k_{\max})
		 \right.\right.
		 \nonumber \\
		 & & \phantom{\int^{k_{\max}+0}_{-k_{\max}-0}E\psi_k|^2
		 \frac{d\psi_k}{dk}}
		 \left.\left.
		 \mbox{}-\psi_k\delta(k-k_{\max})
		 \right|^2\,
	\right]
	\frac{dk}{2\pi} \; .
	\label{eq:LReg}
\end{eqnarray}

This is infinite unless we require the coefficients of the
$\delta$-functions to vanish.  That is, $\psi_k = 0$ at $k = \pm
k_{\max}$.  The Euler--Lagrange equation is $(d^2/2dk^2 + E)\psi_k =
0$, which has the solutions $\exp \pm i(2E)^{1/2}k$. These waves would
propagate to infinity if it were not for the reflecting boundary
conditions at $\pm k_{\max}$ we have just derived.

That is, we have removed the continuum by box quantization in
$k$-space.  In the following we shall do the same for the ballooning
mode problem.

As in \cite{dewar-glasser83} we write the magnetic field of an 
arbitrary three-dimensional toroidal equilibrium plasma with nested 
magnetic flux surfaces labeled by an arbitrary parameter $s$ as
	\begin{math}
	\Bvec = \grad \zeta \cross \grad\psi - q \grad \theta \cross \grad
	\psi \equiv \grad\alpha\cross\grad\psi
	\end{math},
where $\alpha \equiv \zeta - q\theta$.
Here, $\theta$ and $\zeta$ are the poloidal and toroidal angles,
respectively, $\psi(s)$ is the poloidal flux function,
and $q(s)$ is the inverse of the rotational transform. Since 
$\Bvec\dotv\grad s = \Bvec\dotv\grad \alpha = 0$, $s$ and $\alpha$ serve 
to label an individual field line.

We take the stream function \cite{dewar97c} to be given by
    \begin{math}
        \varphi = \widehat{\varphi} \exp(iS-i\omega t) \;,
        \label{eq:eikonalrep}
    \end{math}
where $\widehat{\varphi}(\theta|s,\alpha)$ is assumed to vary on the
equilibrium scale. The phase variation is taken to be rapid,
so $\kvec \equiv \grad S$ is ordered to be large.  The frequency
$\omega$ is ordered $O(1)$, which requires that the wave vector be
perpendicular to $\Bvec$: $\kvec\dotv\Bvec \equiv 0$.  (In this study we
consider unstable ideal MHD modes, $\omega^2 < 0$.)

It immediately follows that the eikonal is constant on each field line: 
$S = S(\alpha,s)$.  From the definition of the wave vector,
    \begin{math}
        \kvec
        = k_{\alpha}\grad\alpha + k_s\grad s
        \equiv k_{\alpha}[\grad\alpha + \theta_k q'(s)\grad s]
        \label{eq:kvec}
    \end{math}
where $k_{\alpha} \equiv \partial S/\partial \alpha$ and $k_s
\equiv \partial S/\partial s$.  Here the anglelike ballooning
parameter $\theta_k$ appears naturally as the ratio
$k_s/q'(s)k_{\alpha}$ \cite{dewar-manickam-grimm-chance81}.

The ballooning equation emerges in the large $|\kvec|$ expansion
\cite{dewar-glasser83,dewar97c} as an ordinary differential equation
to be solved on each field line $(\alpha,s)$ with given
$(k_{\alpha},k_s)$ under the boundary condition
$\widehat{\varphi}(\theta) \rightarrow 0$ at infinity to give the
eigenvalue $\lambda(\alpha,s,k_{\alpha},k_s)$.  This constitutes a
local dispersion relation $\lambda \equiv \rho\omega^2$ (the mass
density $\rho$ being assumed constant everywhere).

The ray equations are the characteristics of the eikonal
equation
$\lambda(\alpha,s,\partial_{\alpha}S,\partial_s S) = \rho\omega^2$. 
These are Hamiltonian equations of motion with $\alpha,s$ the
generalized coordinates, $k_{\alpha},k_s$ the canonically conjugate
momenta, and $\lambda$ as the Hamiltonian.

In axi- or helically symmetric systems all field lines on a given
magnetic surface are equivalent---$\alpha$ is ignorable and
$k_{\alpha}$ is a constant of the motion.  In this case the equations
are integrable and semiclassical quantization can
be used to predict the approximate spectrum of global ballooning
instabilities \cite{dewar-manickam-grimm-chance81}.  This technique
can sometimes be applied successfully, even in nonsymmetric
systems, if there are regions of phase space with a large measure of
invariant tori \cite{nevins-pearlstein88,cooper-singleton-dewar96}. 
In \cite{cooper-singleton-dewar96} this was verified using the global 
eigenvalue code TERPSICHORE \cite{anderson_etal90}.

At the other extreme, if the ray orbits are chaotic (but still
bounded) then the global spectrum is not regularly structured, but
must rather be described statistically by the density of states and
the probability distribution of level spacings using the techniques of
quantum chaos theory (see e.g. \cite{gutzwiller90,ott93}).

However, because of the scale invariance of the ideal MHD equations,
$\lambda$ depends only on the direction of $\kvec$, \emph{not} on its
magnitude: $\lambda = \lambda(\alpha,s,\theta_k)$.  This has the
consequence that the ray orbits are \emph{unbounded} in phase space,
so, strictly speaking, ideal MHD gives rise to a quantum \emph{chaotic
scattering} \cite{gutzwiller90,ott93} problem rather than a straight
quantum \emph{chaos} problem.  This leads to the continuous spectrum
\cite{dewar-glasser83} with singular generalized eigenfunctions that
cannot really be represented using the simple eikonal ansatz.

On the other hand, the absence of a natural length scale in ideal MHD
is a mathematical artifact.  Physically, the ion Larmor radius
provides a lower cutoff in space, or an upper cutoff in $|\kvec|$,
beyond which ideal MHD ceases to apply.  The ballooning equation is
also physically regularized by inclusion of diamagnetic drift
\cite{tang-dewar-manickam82,nevins-pearlstein88}.

However, since in general it leads to a complex ray
tracing problem \cite{hastie-catto-ramos00}, we shall not attempt to model
diamagnetic drift stabilization in this paper.  Rather, we regularize
the ray equations simply by adding a barrier term to the effective
ray ``Hamiltonian'' $H(\alpha,s,k_{\alpha},k_s)$,
\begin{equation}
	H = \lambda(\alpha,s,k_{\alpha},k_s) + U(k_{\alpha}) \;,
	\label{eq:H}
\end{equation}
where the barrier potential we use is $U(k_{\alpha}) \equiv
K(|k_{\alpha}|-k_{\mathrm{max}})^2$ for $|k_{\alpha}| >
k_{\mathrm{max}}$ and 0 for $|k_{\alpha}| < k_{\mathrm{max}}$.  In the
limit of the constant $K \rightarrow \infty$, this infinite box
potential gives the ideal MHD ray equations for $|k_{\alpha}| <
k_{\mathrm{max}}$ and reflecting boundary conditions at $|k_{\alpha}|
= k_{\mathrm{max}}$.  Thus we have a two-degree of freedom Hamiltonian
billiard problem.

Although overly crude for modeling FLR stabilization, the cutoff at
$|k_{\alpha}| = k_{\mathrm{max}}$ provides a reasonable model for
representing the finite spectral bandwidth in the toroidal Fourier
mode number ($n$) representation used in the global eigenvalue codes
TERPSICHORE \cite{anderson_etal90} and CAS3D \cite{schwab93}.

\begin{figure}[tbp]
	\centering
$\begin{array}{cc}
	\includegraphics[scale=0.5]{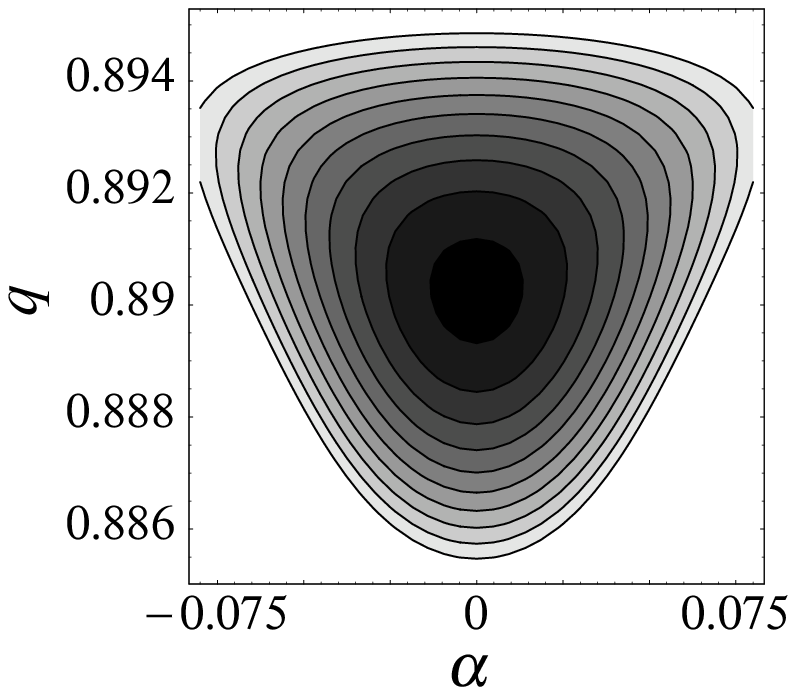} &
	\includegraphics[scale=0.5]{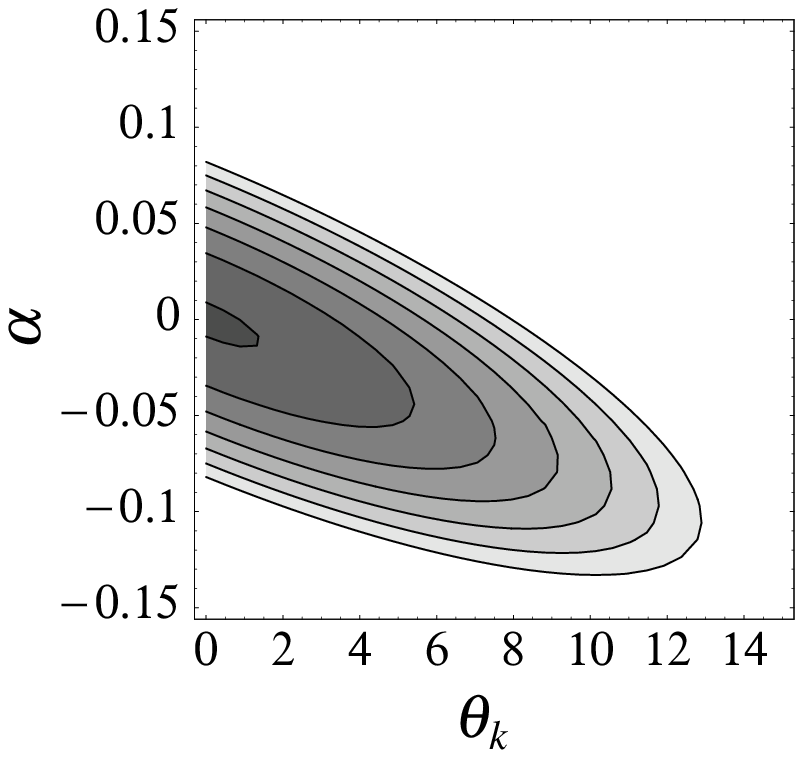}
	\end{array}$
	\caption{The sections $\theta_k=0$ and $q=0.893$ of
	the topologically spherical isosurfaces of the central, (0,0),
	ballooning mode branch, bounded by the isosurface $\lambda = -6$
	(arbitrary  units).  The darker shades denote higher growth rates,
	the peak corresponding to $\lambda \approx -8$.}
	\label{fig:contours}
\end{figure}

Using ballooning-unstable plasma equilibria calculated for the H-1NF
heliac \cite{hamberger_etal90,blackwell01} using the VMEC code
\cite{hirshman-betancourt91}, detailed parameter scans have been
undertaken for two cases.  The first case studied
\cite{cuthbert-dewar00} was obtained by increasing the pressure
gradient of a marginally stable equilibrium \cite{cooper-gardner94}
uniformly across the plasma and thus was ballooning unstable at the
edge of the plasma.  The ray tracing problem for this case would
involve consideration of the effect of the plasma boundary.

Thus a second equilibrium, ballooning stable near the edge of the
plasma, was calculated for the purposes of the present paper.  This
case has a more peaked pressure profile than the first, but both have
average $\beta \approx 1\%$, where $\beta$ is the ratio of plasma 
pressure to magnetic field pressure.

The $q$-profiles are not monotonic---in the peaked pressure profile
case studied in this paper, $q$ was 0.8895 on the magnetic axis,
rising to a maximum value of 0.8964 quite close to the magnetic axis,
then falling monotonically to 0.8675.  Clearly the (global) magnetic
shear is very weak.  Despite this fact and the non-monotonicity, there
is some formal simplification in choosing $s \equiv q$, and we have
taken $s = q$ since the region of plasma studied is in a
monotonic-decreasing part of the $q$-profile (the decreasing region 
outside the maximum-$q$ surface).

In these scans the most unstable ballooning eigenvalue was tabulated
on a three-dimensional grid in $s, \alpha, \theta_k$ space.  The
dependence on $\alpha$ was found to be rapid.  The dependence on
$\theta_k$ was much slower, but the variation was sufficient that the
higher-growth-rate isosurfaces formed a set of distinct, topologically
spherical branches.  It was argued in \cite{cuthbert-dewar00} that
this branch structure is produced by Anderson localization in bad
curvature regions due to the strong breaking of both helical and
axisymmetry in H-1NF.

According to the perturbation expansion in $q'$ described in
\cite{cuthbert-dewar00}, a quadratic form in $\alpha,\theta_k$ should
form a good approximation to $\lambda-\lambda_{\mathrm{min}}(q)$ in
the neighborhood of the central branch.  Accordingly a least-squares
fit on each surface was performed to provide a simple analytical
description of the $(0,0)$ \cite{cuthbert-dewar00} branch.

The radial dependence of the fitting coefficients was 
approximated by fitting to third-degree polynomials in $q$. Sections 
of the resulting approximation to the central branch are shown in 
\fig{fig:contours}. The isosurface spans a substantial
range of magnetic surfaces within the plasma --- the narrow range
of variation in $q$ is due to the low magnetic shear in H-1NF.

In order to establish the nature of the ray dynamics described by the
regularized Hamiltonian, \eq{eq:H}, a numerical integration with
cutoff at $k_{\mathrm{max}} = 50$ was performed with initial
conditions $q=q_2$, $\alpha=0$, and $k_{\alpha}=5$, where $[q_1,q_2] =
[0.8852,0.8951]$ is the $q$-range spanned by the $\lambda=-6$
isosurface as seen in \fig{fig:contours}.  (A run with $k_{\alpha}=10$
was also performed, with similar results.)  Choosing the value $K=1$
gave a good compromise between the sharp boundary potential to be
modeled, and the smooth potential required for the numerical
integration.  The orbit remained on the ``energy shell'' $\lambda =
-6$ to within an accuracy of one part in $10^6$ over the ``time''
interval of the integration, $7500$.
\begin{figure}[tbp]
	\centering
$\begin{array}{cc}
	\includegraphics[scale=0.5]{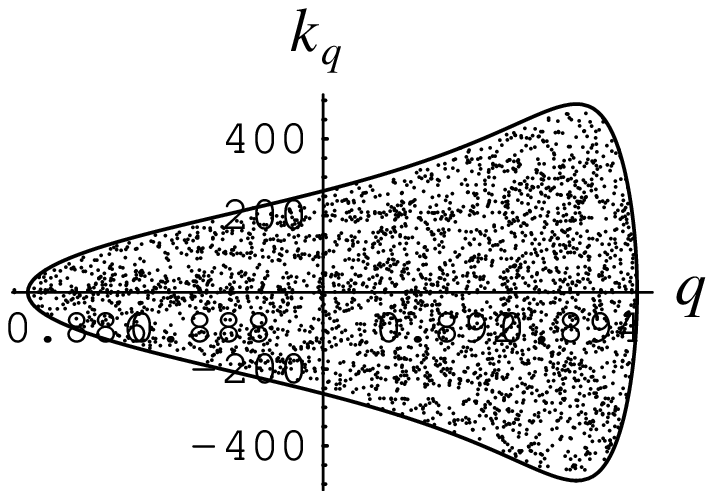} &
	\includegraphics[scale=0.5]{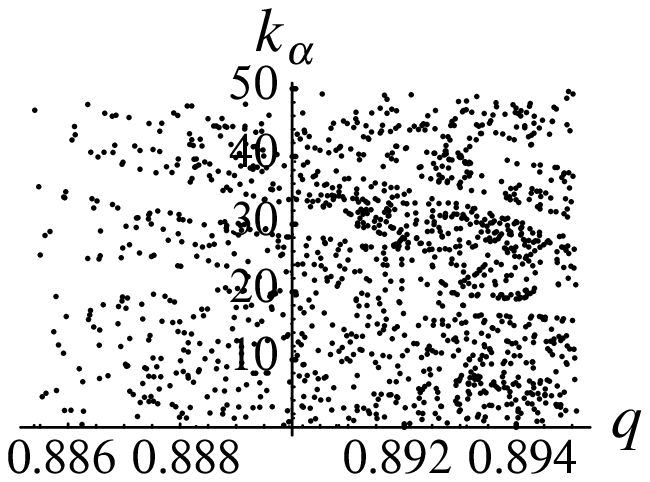}
\end{array}$
	\caption{Two views of 
	intersections with the Poincar\'{e} surface of section $\alpha = 0$.}
	\label{fig:Poincare}
\end{figure}

The two Poincar\'{e} plots in \fig{fig:Poincare} show the orbit to be
strongly chaotic, filling the ``energy shell'' ergodically, except
that the regions $k_{\alpha}>0$ and $k_{\alpha}<0$ are dynamically
disjoint.  The solid curve shown surrounding the outer limits of the
``energetically accessible'' region is calculated by solving
$\lambda(0,q,k_q/k_{\mathrm{max}}) = -6$.

According to the Weyl formula \cite[pp.  257--261]{gutzwiller90}, the
number, $N(\lambda_{\mathrm{max}})$, of global eigenmodes with
eigenvalues below the eigenvalue $\lambda_{\mathrm{max}}$ is given,
asymptotically in the limit $N \rightarrow \infty$, as
$N(\lambda_{\mathrm{max}}) \sim
v_{4\mathrm{D}}(\lambda_{\mathrm{max}})/(2\pi)^2$.  Here
$v_{4\mathrm{D}}(\lambda_{\mathrm{max}})$ is the volume of the
dynamically acessible 4-dimensional phase-space region
$\lambda(\alpha,q,k_q/k_{\alpha}) < \lambda_{\mathrm{max}}$,
$0<k_{\alpha}<k_{\mathrm{max}}$. The $k_{\alpha}$ integration can be 
performed analytically, giving
\begin{math}
	v_{4\mathrm{D}}(\lambda_{\mathrm{max}})
	= \frac{1}{2} k_{\mathrm{max}}^2 
	v_{3\mathrm{D}}(\lambda_{\mathrm{max}}),
\end{math}
where $v_{3\mathrm{D}}(\lambda_{\mathrm{max}})$ is the volume within
the isosurface $\lambda(\alpha,q,\theta_k) = \lambda_{\mathrm{max}}$. 
Thus
\begin{equation}
	N(\lambda_{\mathrm{max}})
	\sim \frac{1}{8\pi^2} k_{\mathrm{max}}^2 
	v_{3\mathrm{D}}(\lambda_{\mathrm{max}}) \;.
	\label{eq:Nstates}
\end{equation}

We can make a rather rough estimate of the minimum value of
$n_{\mathrm{max}}$ required for CAS3D or TERPSICHORE to find even one
eigenvalue with $\lambda < \lambda_{\mathrm{max}}$ by setting
$N(\lambda_{\mathrm{max}}) = 1$ and calculating $k_{\mathrm{max}}
\approx n_{\mathrm{max}}$ from \eq{eq:Nstates}.  This gives
$n_{\mathrm{max}}(N=1) \sim (8\pi^2/v_{3\mathrm{D}})^{1/2}$.

The isosurface $\lambda = -6$ studied above is about the largest of
the disjoint topologically spherical isosurfaces corresponding to the
highly toroidally localized strongly ballooning unstable regions of
$\alpha,q,\theta_k$ space.  (For $\lambda > -6$ the isosurfaces are no
longer topologically spherical.)  Using the polynomial fits described
above, we calculate $v_{3\mathrm{D}}(-6) = 0.02158$.  This gives
$n_{\mathrm{max}}(N=1) \approx 60$.  Assuming that the dominant
contributions to the MHD energy $\delta W$ come from the rational
surfaces intersecting the $\lambda = -6$ isosurface, we thus predict
that it would be necessary to include, as a minimum set, basis
functions corresponding to one of the two ``mode families''
\cite{schwab93} contained in the set $(n,m) = (9,8)$, $(18,16)$,
$(19,17)$, $(27,24)$, $(28,25)$, $(35,31)$, $(36,32)$, $(37,33)$,
$(38,34)$, $(44,39)$, $(45,40)$, $(46,41)$, $(47,42)$, $(53,47)$,
$(54,48)$, $(55,49,)$, $(56,50)$, and $(57,51)$ to resolve a
toroidally localized ballooning mode.  (Here $n,m$ are the toroidal
and poloidal Fourier mode numbers, respectively.)

The large value of $n_{\mathrm{max}}(N=1)$ required, and the unusual
spread in $n$ required in the basis set, will make these modes difficult to
resolve using global eigenvalue codes (e.g. the simplifying phase
factor method sometimes used in CAS3D studies \cite{reiman_etal97}
would not be appropriate).  It is hoped that the Weyl formula estimate
above will act as a guide in a future more extensive study using such
a code.  Physically, the large value of $n_{\mathrm{max}}$ suggests
that toroidally localized ballooning modes in H-1NF should be subject
to strong FLR stabilization.

We can also apply the same approach to the toroidally localized ballooning 
branches found in the Large Helical Device (LHD) study \cite{cuthbert_etal98}.
From the plots in \cite{cuthbert_etal98} we estimate $v_{3\mathrm{D}} 
\sim 0.05$, which gives $n_{\mathrm{max}}(N=1) \approx 40$.

The ballooning calculations were carried out on the Australian
National University Supercomputer Facility's Fujitsu VPP300 vector
processor.  We thank Dr.\ H.~J. Gardner for providing the H-1 heliac
VMEC input files and Dr.\ S.~P. Hirshman for use of the VMEC
equilibrium code.  Some of this work was done while one of us (RLD)
was a visiting scientist at Princeton University Plasma Physics
Laboratory, supported under US DOE contract No.  DE-AC02-76CH0-3703. 
Useful conversations with Drs.  M. Redi and A.H. Boozer are gratefully
acknowledged.

% \begin{figure}[tbp]
% 	\centering
% $\begin{array}{cc}
% 	\includegraphics[scale=1.0]{Contours_aq.eps} &
% 	\includegraphics[scale=1.0]{Contours_tka.eps}
% 	\end{array}$
% 	\caption{The sections $\theta_k=0$ and $q=0.893$ of
% 	the topologically spherical isosurfaces of the central, (0,0),
% 	ballooning mode branch, bounded by the isosurface $\lambda = -6$
% 	(arbitrary  units).  The darker shades denote higher growth rates,
% 	the peak corresponding to $\lambda \approx -8$.}
% 	\label{fig:contours}
% \end{figure}
% 
% \begin{figure}[tbp]
% 	\centering
% $\begin{array}{cc}
% 	\includegraphics[scale=1.0]{Poincare.eps} &
% 	\includegraphics[scale=1.0]{Poincare_qka.eps}
% \end{array}$
% 	\caption{Two views of 
% 	intersections with the Poincar\'{e} surface of section $\alpha = 0$.}
% 	\label{fig:Poincare}
% \end{figure}

\end{document}